\documentclass[a4paper]{sig-alternate-10pt}

\usepackage{graphicx, epsfig}

\DeclareFixedFont{\auacc}{OT1}{phv}{m}{n}{12}
\DeclareFixedFont{\afacc}{OT1}{phv}{m}{n}{10}

\newtheorem{scheme}{Scheme}

\newcommand{\otiy}{{\sffamily Otiy}}

\begin{document}

\conferenceinfo{CoNext}{'2007 Columbia, NY USA}
\CopyrightYear{2007}

\title{Otiy: Locators Tracking Nodes}

\numberofauthors{1}
\author{
\alignauthor
     Mathias Boc, Anne Fladenmuller, and Marcelo Dias de Amorim \vspace{2mm}\\
     \affaddr{Laboratoire d'Informatique de Paris 6 (LIP6/CNRS)}\thanks{This work has been partially
supported by the IST European project WIP under contract 27402 and
by the RNRT project Airnet under contract 01205.}\\
     \affaddr{Universit{\afacc\'e} Pierre et Marie Curie~-- Paris 6}\\
     \email{\{boc,fladenmu,amorim\}@rp.lip6.fr}
}

\maketitle

\begin{abstract}
We propose \otiy, a node-centric location service that limits the
impact of location updates generate by mobile nodes in IEEE~802.11-based
wireless mesh networks. Existing location services use node
identifiers to determine the locator (aka anchor) that is
responsible for keeping track of a node's location. Such a strategy
can be inefficient because: (i) identifiers give no clue on the
node's mobility and (ii) locators can be far from the
source/destination shortest path, which increases both location
delays and bandwidth consumption. To solve these issues, \otiy\
introduces a new strategy that identifies nodes to play the role of
locators based on the likelihood of a destination to be close to
these nodes~-- i.e., locators are identified depending on the
mobility pattern of nodes. \otiy\ relies on the cyclic mobility
patterns of nodes and creates a slotted agenda composed of a set of
predicted locations, defined according to the past and present
patterns of mobility. Correspondent nodes fetch this agenda only
once and use it as a reference for identifying which locators are
responsible for the node at different points in time. Over a period
of about one year, the weekly proportion of nodes having at least
50\% of exact location predictions is in average about 75\%. This
proportion increases by 10\% when nodes also consider their closeness
to the locator from only what they know about the network.
\end{abstract}

\section{Introduction}
\label{sec:introduction}

Promoting node mobility in self-organizing wireless networks implies
setting up an efficient location management scheme. Indeed, the
degree of node mobility impacts the amount of location updates
disseminated throughout the network; this problem becomes even more
critical in dense areas, where the profusion of signaling messages
induces serious contentions and penalizes the overall performance of
applications.

There are different techniques to implement location services. In
flooding-based approaches, no proactive decision is made; when a
source wants to communicate with a destination, it floods the entire
network with a lookup message. Although simple, such an approach is
very resource consuming and thus inappropriate for wireless
networks. In order to solve this problem, some solutions propose to
use location anchors~-- when a source wants to communicate with a
destination, it must first ask the anchor about the current location
of the node. Anchor-based architectures can be implemented in a
centralized or distributed
manner~\cite{mobile-slalom,hubaux01towards,stojmenovic99routing,
woo01scalable}. Nevertheless, these solutions share a common
pitfall: {\it they have no control on the location of the anchor}.
This means that a source may have to send a lookup message to a
far-away anchor for a destination that is possibly nearby. The
consequences are twofold: (i) lookup phase may experience large
delays and (ii) lookup messages may travel long distances, reducing
the overall capacity of the network. The consequences are the same
for the destination node which also have to update its current
location to a possibly far-away responsible (or assigned) anchor.

To address the abovementioned problems, we propose
\otiy\footnote{Oti'y is a Creole word meaning ``where can I find
him?''}, a two-tiered location service that relies on the fact that
most nodes present cyclic mobility patterns. Indeed, many works have
shown that, in many situations, nodes do show mobility
characteristics that can be quite well predicted based on the node's
mobility history~\cite{castro_charac,ghosh_profiling,francois,
wu01personal,spatial_temporal}.

\otiy\ benefits from mobility prediction by decoupling the location
service into two tiers as follows:

\begin{enumerate}

\item {\bf Global service.} This service determines the {\it anchor}
of a node, in a similar way to traditional solutions. The difference
here is that anchors in \otiy\ do not store the current location of
nodes; instead, it returns an {\it agenda} containing for each
period of the day the location server (or {\it locator}, cf., next
bullet) which is most probably the closest to the mobile node. This
agenda is available for a reasonable amount of time and thus can be
stored by the communicating nodes to prevent frequent access to the
global service.

\item {\bf Local service.} Locators are points in the infrastructure
that effectively know at a given time where mobile nodes are. They respond to lookup
requests and inform about the current position of mobile nodes. To
determine which of its locators is currently in charge of storing
its location, a mobile node refers to its agenda. As a locator is
chosen by a  mobile node because of its high probability of being
close to it, location  updates remain localized.

\end{enumerate}

The idea behind this system is to use the global service from time
to time and the local service most of the time~-- and thus reduce
the overhead found in flat solutions. More specifically, an agenda
contains a list of  pairs ${\tt (Time\_slot, Locator)}$, where for
the time period ${\tt Time\_slot}$ the location information of the
mobile node will be managed by ${\tt Locator}$. This agenda remains
valid for a certain period, after which it must be renewed at the
corresponding $anchor$. A source willing to communicate with a
destination first fetches the agenda of the destination at its
corresponding anchor. Then, during the validity period of the
agenda~-- which can last for weeks depending on the prediction
accuracy~-- the source only contacts locators to obtain the exact
location of the destination. The generation of a pertinent agenda to
identify the best locator (closest to the mobile node) for each time
slot is thus at the heart of \otiy. We argue that mobile nodes don't
necessarily have a positioning system and should therefore determine
their mobility pattern based only on topological information,
meaning the logs of attachment to the access points.

Although the basic concepts of \otiy\ can be generalized to
different types of self-organizing wireless networks, in this paper
we focus on the context of wireless mesh networks (WMN) composed of
IEEE~802.11 access points (mesh routers). In this case, mobility is
defined as a sequence of access points a node associates to along
time. We will see later in this paper that defining exact mobility
patterns in a wireless mesh network is a complex task. For instance,
misinterpretations of mobility may happen mainly due to ping-pong
effects, which  are oscillations of associations/disassociations to
nearby mesh routers due to changes in medium conditions. In order
to address this problem, we introduce a self-organizing scheme that
groups nearby mesh routers into clusters from each node standpoint.
This clustering scheme masks ping-pong effects and allows reducing
the space of possible locations.

We evaluate \otiy\ under a large population of nodes using real
traces of mobility in a campus
scenario~\cite{dartmouth-campus-syslog}. We find that nodes,
although heterogeneous in nature, do show cyclic behaviors according
to their own rules. The analysis covers periods ranging from one
month to more than one year and presents results in function of 
the number of active nodes and the different periods animating 
the campus (e.g., holidays, school periods, weekends). From the 
observations, the weekly proportion of nodes having at least 
50\% of exact location predictions is in average about 75\%. 
This proportion increase by 10\% if nodes also consider known 
properties on the deployment area (e.g., buildings, offices, paths).

In Section~\ref{sec:basics}, we present the rationale for a
node-centric approach and introduce \otiy's architecture. In
Section~\ref{sec:clustering}, we present our algorithm to readapt
the association logs and to elect the appropriate locators. In
Section~\ref{sec:cyclic}, we analyze the cyclicity and the
persistence of the nodes' behaviors upon which is based \otiy. In
Section~\ref{sec:evaluation}, we evaluate the agendas accuracy 
according to the wireless data traces from the Dartmouth
campus. We delay our discussion of related work until
Section~\ref{sec:relatedwork} in order to have enough context to
make the necessary connections. We finally present some conclusion
in Section~\ref{sec:conclusion}.

\section{Otiy's design}
\label{sec:basics}

\otiy\ introduces a different approach for distributed location
services in wireless mesh networks. In this section we first
motivate our proposal and then present \otiy's architecture and
operation. For lack of space, many details are omitted.

\subsection{Rationale}
\label{subsec:rationale}

We consider wireless mesh networks with following characteristics:
(i) resources are scarce (wireless me\-dium), (ii) the backbone is
static, (iii) nodes are mobile, and (iv) the backbone can be highly
dense (e.g., for over-provisioning) in localized hotspot areas. In
such a context, which is expected to happen in many situations, the
networking architecture becomes fully dependent on an {\it efficient
location service}. An ``efficient'' location service should have
{\it at least} the following characteristics:

\begin{enumerate}

\item Location updates do not interfere much with 
the network operations.

\item Location signaling messages have low latency.

\item Location information is accurate.

\end{enumerate}

This paper provides a response to these three requirements. We base
our reasoning on the possibility of having {\it persistent location
information} (i.e., with long validity duration)~-- location updates
to the anchor (called location dissemination in the rest of the paper) 
become then more spaced in time, which reduces the amount of propagated 
signaling messages. This is the response to requirement~\#1 above.

To provide persistent location information, we take as a premise
that different nodes have different levels of mobility. Many studies
in the literature have shown that there is a large part of
predictability in how nodes move at an
AP~\cite{francois,song_predict,franc_predict} granularity. Location
services that are based on these studies use prediction to precisely
identify the APs to which nodes will be associated with.
Nevertheless, the presence of the ping-pong effect and the
development of cognitive radio can have major impact on the
efficiency of these approaches.

In \otiy, we take a different approach. Instead of trying to obtain
the exact location of a node, we use the node's past and present
mobility pattern to distribute locators throughout the network. With
high probability (as shown in Section~\ref{sec:cyclic}), locator
nodes are placed close to the current location of the node and so,
close to the shortest path between sources and destinations. This
responds to both requirements~\#2 and~\#3.

\subsection{The reasons for node-centric mobility}
\label{subsec:nodecentric}

We have identified four aspects related to the mobility of nodes:

\begin{itemize}

\item \textbf{Periods of activity.} Depending on the device type and
the necessity to being connected, nodes can show completely different 
periods of activity, ranging from diurnal activity to sporadic connections 
only on week-ends.

\item \textbf{Mobility coverage.} We refer to ``mobility
coverage'' as the number of different APs visited during a single
session. We observe that some nodes are aware of the roaming
capabilities offered by their underlying network and take advantage
of them, while others remain static or have only a nomadic behavior.

\item \textbf{Home location characterization.} It is not always
possible to identify for each node a home location (an area where a
node spends more than 50\% of its association
time~\cite{henders_usage}). This parameter gives an indication on
the regularity of the associations of the nodes to a specific
location.

\item \textbf{Number of visited areas.} This parameter is
an extended view of the mobility coverage; the number of visited
areas accounts for all visits of a node for the entire observed
period (multiple sessions). Mobile or not, the disparities between
nodes on this point are deep. The interest of visiting different APs
in the network has clear relationship with the social interest of
visiting different areas in the environment.

\end{itemize}

Interests, constraints, and motivations behind the pat\-terns of
mobility are manifold and have different impact on the complexity of
network management. Further\-more, each node has its own mobility
pattern, which itself varies in time. For these reasons, we advo\-cate
that, for a location service to be efficient, it must manage
mobility at a node scale.

\otiy\ goes a bit further and proposes that nodes self-profile their
behaviors. Such a node-centric approach consists in making nodes
themselves log their sequence of associations together with
timestamps and SSIDs. We show in the following how \otiy\ makes use
of such information.

\subsection{Agenda of locators}
\label{subsec:agenda}

The key element of our proposal is the agenda of locators. The image
of the agenda is important and contributes to highlight the
relationship we want to give between a precise location and the
considered time.

We delay our discussion of cyclicity of mobility until
Section~\ref{sec:cyclic} in order to avoid interrupting our
reasoning. In this way, we ask the reader to just assume for the
moment that nodes have cyclic mobility.

\subsubsection{Description}
\label{subsubsec:description}

We define the agenda for node $n$ as table $A_{n}$. It is composed
of $N$ columns defining the days of cycle and $M$ rows that give the
granularity of the estimation we give to the node's mobility.
W.l.g., in this paper we consider a cycle of one week and a
granularity of one hour (i.e., $N=7$ and $M=24$). The values of $N$
and $M$ will be motivated in Section~\ref{sec:cyclic}). We also
assign a validity period for the agenda, which is a multiple of $N$
representing the number of full cycles the agenda will remain
unchanged (no updates will be provided).

To each time slot $a_{ij}$ (with $0 \leq i<N$ and $0\leq j<M$) of an
equal duration $D=\lceil\frac{24}{M}\rceil$ hour(s), is assigned one
locator node. The choice of this locator depends on the mobility of
the node observed during the same time slot in the past and is
typically a mesh router network address.

We store the history of mobility pattern of node $n$ in a set of
tables $H^k_{n}$ (of same dimensions as $A_{n}$), where $k \geq 0$
indicates the ``age'' of the table. We define $H_{n}^0$ as the table
containing the current mobility of the node, $H_{n}^1$ the mobility of
the precedent week and so on until $k=k_{\max}$ ($k_{\max}$
typically varies between 2 and 4). In each time slot $h_{i,j}^k$,
the node records the location of the area at which the node spent
most of its time during slot $j$ for the day $i$, $k$ weeks before,
according to the mobility log file. In this way, we define a notion
of prevalence of a particular area for a given time slot.
Furthermore, for each $h_{i,j}^0$ the node records the duration of
association $d(h_{i,j}^0)$ (where $0 < d(h_{i,j}^0) \leq D$) of the
selected location.

\begin{figure*}[t]
 \centering
 \includegraphics[width=12cm]{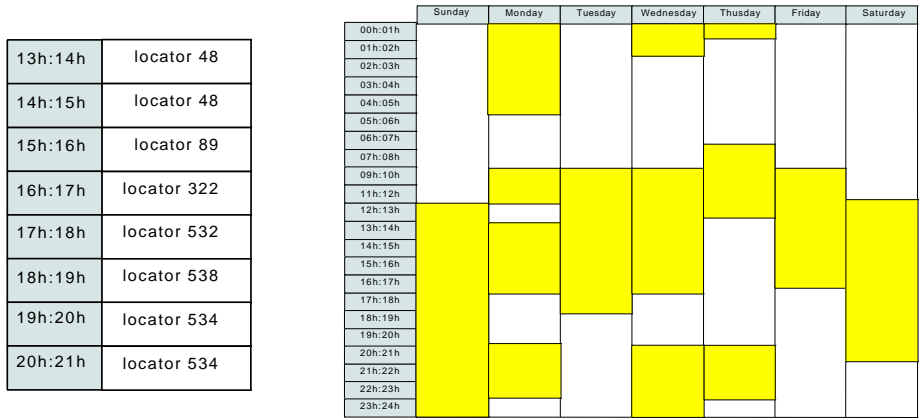}
 \caption{Example of a simplified agenda of associations.}
\label{fig:description}
\end{figure*}

At the end of each validity period, the node generates a new agenda
based on its mobility history (as on Fig.~\ref{fig:description}).
The resulting locator $a_{i,j}$ will depends on the oldness he have
in the network, the necessary mobility history needed to provide an
accurate location $k_{\max}$ (studied in
Section~\ref{sec:evaluation}), and the duration of association
$d(h_{i,j}^0)$.

We can now define the locator $a_{i,j}$ as:

\begin{equation}
a_{i,j}= \left\{
          \begin{array}{ll}
        h_{i,j}^0 & \qquad \mathrm{if}\quad d(h_{i,j}^0)=D\\
        \text{maxoccur}(h_{i,j}^0,...,h_{i,j}^{k_{max}}) & \qquad \mathrm{otherwise,}\\
      \end{array}
     \right.
\end{equation}

In the case where $d(h_{i,j}^0)=D$, we assume that the duration of
association at this particular area has been greater than $D$. If in
the node's history it has always been the same area, then giving a
higher weight to the latest observation does not improves the
quality of the estimation. If it is a different area, we conclude
that it indicates a deep change in the behavior of the node and then
we select this locator for the agenda.

If $d(h_{i,j}^0) < D$, we choose the locator which has the maximum
number of occurrences in the mobility history for the same time slot
(if equal, we choose the most recent locator). In this case, we can
not judge on the persistence that $h_{i,j}^0$ will have in the
future; in this way, we give more weight to the node's habits for
this time slot.

\begin{scheme}
To avoid holes in the agenda, we extend the locator of the preceding
non empty slots to cover an empty time slot.
\end{scheme}

\subsubsection{Bootstrap}
\label{subsubsec:bootstrap}

For the first association in the network, the node $n$ generates an
agenda $A_n$ where all $a_{i,j}$ are set to the location of the first
visited mesh router. This agenda has a validity period until the end
of the cycle (in our case the end of the week). At the end of this 
period, the new generated agenda will have logically the values 
given by $H_{n}^0$. After a longer period, the choice of the locators 
become more accurate because the node disposes of a larger history.

\begin{scheme}
The bootstrap procedure is executed for the first association in the
network and also after long periods of inactivity. In this latter
case, we observe that the behaviors of the nodes often become
completely different (e.g., change of home location and different
mobility pattern). The minimal duration of inactivity before a
restart of the bootstrap is discussed in
Section~\ref{sec:evaluation}.
\end{scheme}

\subsection{Location updates}

\begin{figure}
 \centering
 \includegraphics[width=8.2cm, height=5.6cm]{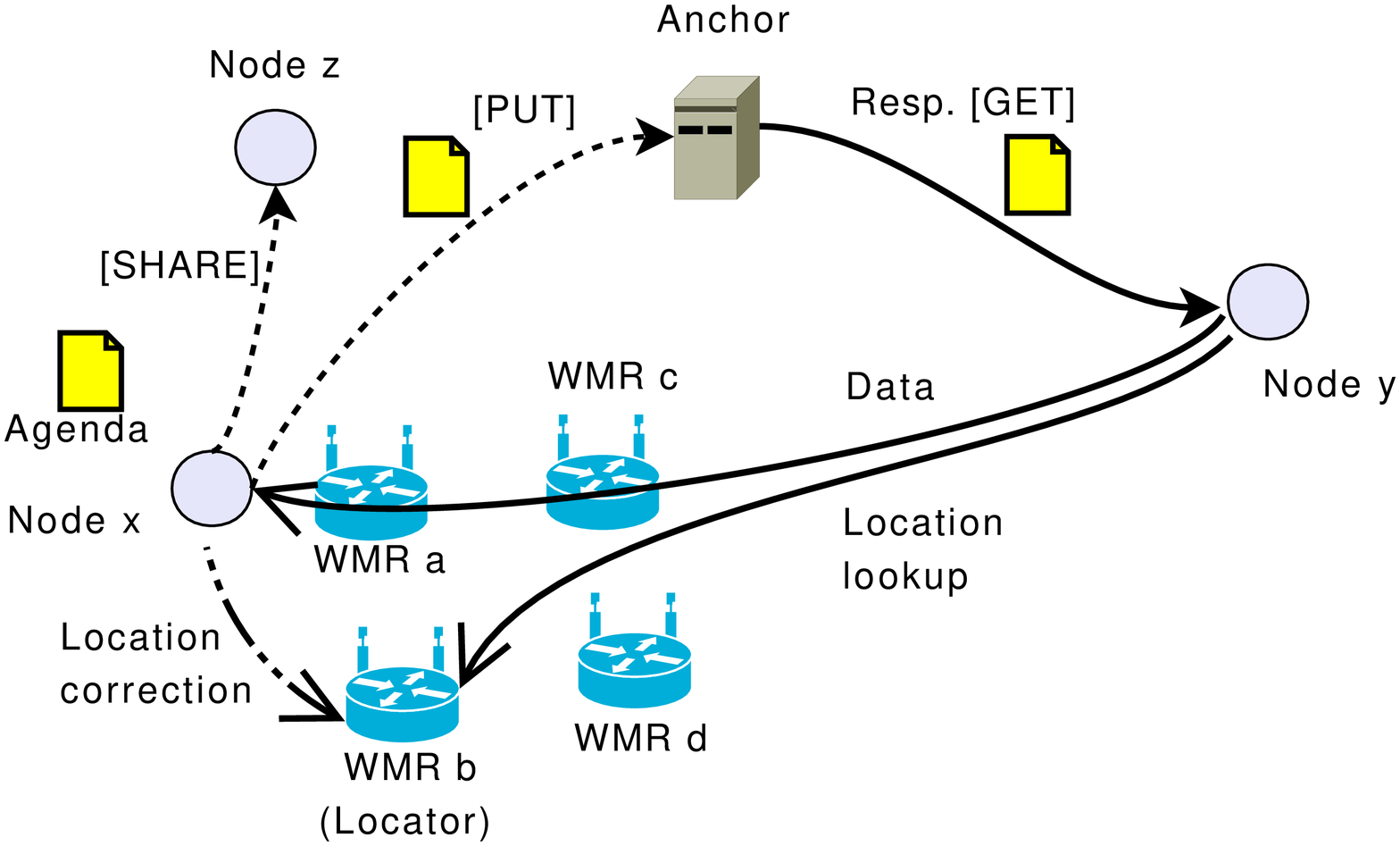}
 \caption{Mechanisms for location dissemination and updates.}
\label{fig:dissemination}
\end{figure}

We make a logical distinction between updating the generated agenda
(called ``location dissemination'') and updating the current
location to the corresponding locator in a given time slot (called
``location correction''). The term ``correction'' is related to the
idea that we hope that the locator will be the point where the node
will be directly associated with.

\subsubsection{Agenda dissemination}

As shown in Fig.~\ref{fig:dissemination}, after having generated the
agenda of locators, the node has two ways to disseminate it (dashed
lines). The first is with the basic primitive of sending (``PUT'') a
copy of the agenda to the responsible anchor. The agenda will thus
be available for every new contact. The second is through a new
primitive ``SHARE''. The node, in a \textsc{P2P} way, can share its
agenda or the agenda of a known contact with others peers. We added
it for two main objectives: (i) to support viral dissemination of
this agenda in order to push decentralization to its limits, and
(ii) to give a community dimension to this agenda. Nodes in the same
social and/or physical community will be able to exchange between
them the agenda of a particular server/device or of a common
friend.\footnote{Although this is an important optimization aspect
of \otiy, we do not address it in details here.}

\subsubsection{Location correction}

There are two situations where location corrections occur
(semi-dashed line in Fig.~\ref{fig:dissemination}):

\begin{itemize}

\item If at $h_{i,j}^0$ the node is still in the area of $a_{i,j-1}$
(or $a_{i-1,M}$ if $j=0$) or on its way to $a_{i,j}$. In this 
cases the node registers its current location to the right locator 
(given by the agenda).

\item If at $h_{i,j}^0$ the node is in the same region than $a_{i,j}$
but not directly associated with the locator.

\end{itemize}

We make this distinction between the two situations because, for the
latter one, one could setup a local mobility management system in
order to limit the number of location corrections. Location
corrections are made pro-actively and are used by the locator to
return the current location of the node when required by
correspondent nodes.

\section{Persistence of Nodes' behavior}
\label{sec:environment}

In \otiy\, each node creates its own agenda, based on simple
information collected from its preceding associations and movements.
In order to validate the concept of agenda, we first need in this
section to validate the assumption that nodes' behaviors are
persistent and mostly periodic by nature. We use the node movements
collected on the wireless access network of Dartmouth
campus~\cite{dartmouth-campus-movement-2005-03-08} to study the
perception nodes can get of their own mobility.

\subsection{Retrieving behaviors from data logs}

\subsubsection{Experimental Data Set}

The data set we use represent three years (2001-04-11 to 2004-06-30)
of collected information about all the wireless cards connecting to
the wireless access network of Dartmouth
campus~\cite{dartmouth-campus-syslog}. The campus is
composed of 188 buildings covered by 566 official APs on 200 acres
and about 5,500 students.

To better understand nodes' mobility characteristics, we focus our
analysis on the movement files accompanying these data
traces~\cite{dartmouth-campus-movement-2005-03-08}. These files
detail the associations and disconnections periods of each
anonymized wireless adapter to any of the APs.  A disconnection is
recorded either as the result of disassociation requests or after 20
minutes of inactivity.

Our analysis is based on a four-week period, from the 5th of January
2004 until the end of the 1st of February of the same year. The
choice of a four-week period is to smooth the impact of punctual
changes in behaviors. Although this period length might seem too
large or too short to capture a complete behavior for a certain
number of observed nodes, we noticed that this is a good compromise
to study the persistence of behaviors.

The choice of the month of January 2004 has been made for three main
reasons: (i) to avoid the reported bugs in the collection of
\textsc{Syslog} events, (ii) because a large number of the devices
were active during this month (just after Christmas holidays), and
(iii) the year 2004 showed more mature nodes.\footnote{We
consider as ``mature nodes'' the nodes that know the network
well enough so that they make use of this knowledge and change their
displacements in function~\cite{henders_usage}.}

\subsubsection{Identification of movements and positions}
\label{sec:understanding}

The comprehension of node mobility is a tough problem when it
relies on raw measurement data. The wireless nature of the network
with all the variations and their consequences, as well as the
density of the APs in the environment, is reflected as variations in
the observed topology. We can cite at least four types of events
that cause these variations:

\begin{enumerate}

\item {\it Ping-pong effect}. It refers to the
succession of as\-sociations-disas\-sociations between two ore more
APs. It is caused by the closeness of the signal to noise ratio
(SNR) of neighboring APs and/or the aggressiveness of the wireless
card.

\item {\it Localized network problems}. If for
some technical reasons, one AP becomes disabled for a certain amount
of time, the node probably associates with another neighboring
AP.

\item {\it Physical micro-variations}. The physical mobility of
nodes is frequently very localized (about a few meters). In
topological dense areas, these micro-variations result in highly
variable association patterns.

\item {\it Erroneous reproducibility.} There is
a probability that the same physical movement results in different
association patterns.

\end{enumerate}

Concerning this latter point, the repetition of movements can create
junctures between those different patterns and create sectors of
micro-mobility which can be detected with a topological standpoint.
The need of an algorithm to recognize these junctures is then
required to better understand the real objectives of the observed
movements. To do so, we introduce a clustering algorithm that uses
roaming events contained in the data logs to help each node
identifying nearby APs from their mobility point of view. \otiy\
relies on these clusters to provide more accurate predictions of
associations.

\subsection{Individual-based clustering}
\label{sec:clustering} \label{sec:algorithm}

The goal of the clustering algorithm we propose in the following is
to identify ``places'' of association that hide areas of
micro-variations (and thus reduce useless location updates). The
methodology is decoupled in two parts: the collection of network
associations and the clustering procedure.

\subsubsection{Collection of network associations}
\label{subsec:events}

The collection of network events is assured by two data structures.
For each node, the relationship with the $M$ APs in the
network are represented through the roaming matrix ${\bf R}=M\times
M$, where each element $r_{ij} \in {\bf R}$ informs about the number
of cumulated roaming events from AP $i$ to AP $j$. The cumulated
roaming events imply that the relationship between two subsets of
APs can appear after independent sessions.

The second data structure is also created in a per-node basis.
It is an $M$-row table that stores general information about each
AP, such as the total number of associations, the cumulated
association duration, and the average association duration.

\subsubsection{The clustering algorithm}
\label{subsec:algorithm}

We now define the terms which will be used in the algorithm:

\begin{itemize}

\item {\it Link}. There is a ``link'' $l_{ij}$ between two APs $i$
and $j$ if there are bi-directional roaming events between these two
APs ($r_{ij}\neq 0$ and $r_{ji}\neq 0$).

\item {\it Cost of a link}. The cost of a link or ``distance''
between two APs $i$ and $j$ is equal to $r_{ij}+r_{ji}$. We thus
define the cost of a link $l_{ij}$ as $c_{ij}=c_{ji}=r_{ij}+r_{ji}$.

\item {\it Cluster}: A ``cluster'' is a group (possibly unitary) of
APs. Each AP belongs to only one cluster. Two APs are eligible to be
merged in the same cluster if there exists a link between them.

\item {\it Weight of a cluster}. The weight $w_{i}$ of a cluster
$c_{i}$ is the value of the maximum link cost within the cluster.

\end{itemize}

We start with a graph whose vertices represent the APs of the
network. An edge exists between two vertices if there is a link (as
defined above) between the corresponding APs. In order to limit the
variations of intra-cluster link costs, we define a threshold $k$
(with $0\leq k\leq 1$).

At the beginning, each AP $i$ becomes a cluster $c_i$ of size 1 and
weight $w_i=0$. We consider first the link with the highest value in
the graph. If two APs $i$ (cluster $c_i$) and $j$ (cluster $c_j$)
can merge together (i.e., $c_{ij}\geq k \times \max\{w_i,w_j\}$),
then the weight of the resulting cluster is equal to the highest
value of the links within the cluster, i.e., $w_{c_{i} \cup c_{j}} =
\max\{w_i,w_j\}$.

\begin{figure}[t!]
\begin{center}
\includegraphics[width=8cm,height=5.5cm]{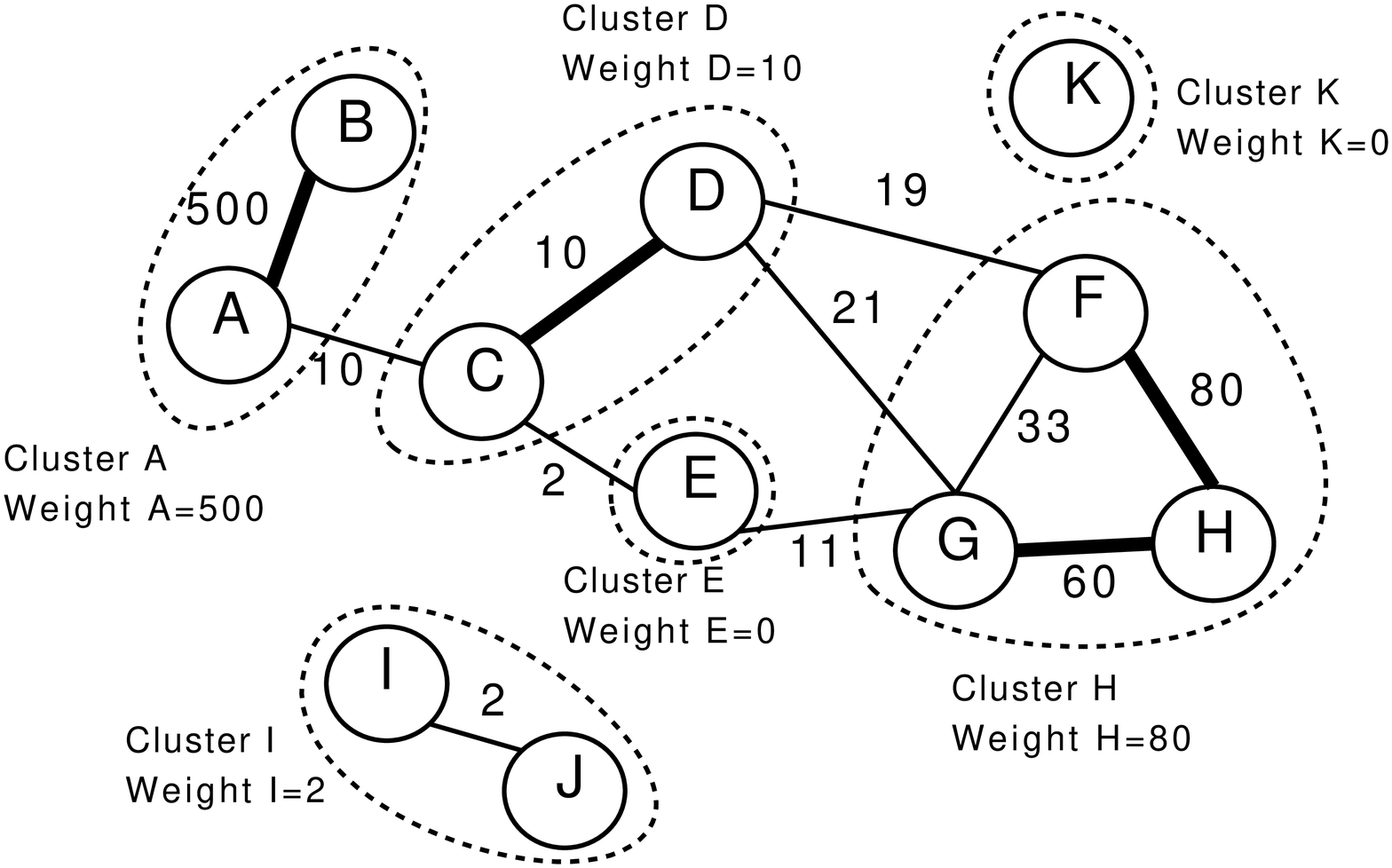}
\caption{Clustering mechanism according to the number of roaming
events between access points (with $k=0.5$).} \label{fig:clusters}
\end{center}
\end{figure}

We repeat the clustering process until there are no more links to be
considered. An example of a resulting clustered graph representation
is illustrated in Fig.~\ref{fig:clusters}. One can notice that the
costs $l_{DF}$ and $l_{DG}$ are not sufficient to merge both
clusters. The same happens with $l_{AC}$, $l_{CE}$, and $l_{EG}$.
Such an approach serves to differentiate paths from locations where
an node stays longer.

We can now define that within a cluster, an eligible locator will 
be the mesh router with the greater cumulated association duration to 
interpret the likelihood to be associated on a particular AP. 

\subsubsection{Resulting properties}

\begin{figure}
\includegraphics[width=8.5cm, height=6.5cm]{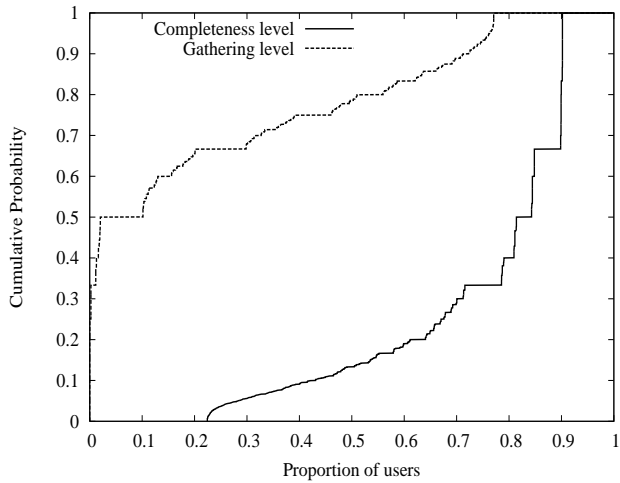}
\caption{CDFs of the completeness level of the resulting embedded
graphs and the gathering level.} \label{fig_clus_properties}
\end{figure}

We study the variations of two important properties of the resulting
embedded graphs generated through our clustering algorithm among the
patterns of mobility of the nodes (see
Fig.~\ref{fig_clus_properties}). We ran the clustering algorithm on
4 weeks by using $k=0.5$ on 4,766 active nodes.\vspace{2mm}

\noindent \textbf{The gathering level.} We define as ``gathering
level'' the ratio of the number of generated clusters on the total
number of visited APs. For 68\% of the nodes, the ratio is
greater than $0.5$ and strictly inferior to 1, while for 23\% the
ratio is equal to $1$. The clustering algorithm thus performs well
as it gathers mostly APs involved in ping-pong effects and high
micro-mobility highlighting the different places of association.\vspace{2mm}

\noindent \textbf{The completeness level.} We define as
``completeness level'' the ratio of the number of edges between the
clusters on the number of edges required to obtain a complete graph.
A graph of $n$ vertices is complete when there is an edge between
every pairs of distinct vertices. This represents $\frac{n(n-1)}{2}$
edges. With this ratio, we have an idea of the closeness of the
different clusters according to the mobility of the nodes.
22\% of the nodes have a ratio equal to $0$. They do not have
inter-clusters mobility and/or have only one generated cluster. In
contrast only 10\% have a ratio equal to $1$ which represents a
limited coverage area of mobility. Finally 58\% of the nodes
have a ratio comprised between $]0,32]$. They have inter-cluster
mobility but not completely connected graphs.

\subsection{Persistence of cyclic behaviors}
\label{sec:cyclic}

\otiy\ relies on notions of persistence of cyclic behaviors and
preferential time of attachment to specific areas. In this section,
we analyze patterns of association with the above-created clusters.
To do so, we rewrite each movement file by replacing APs'
identifiers by their cor\-responding clusters' identifiers and by
aggregating the timestamps of consecutive associations in the same
cluster.

\subsubsection{Cyclic time-related (re)association behavior}
\label{sssec:cyclic_t_ass}

Individuals which have habits in an environment, show cyclic
patterns of mobility at the topological level. We asset this aspect
through the analysis of the re-associations at prevalent locations
in an hourly basis by making distinction between the days of the
week.

For this analysis, we enlarge the observed period to eight weeks to
be sure to not be in a situation where the nodes show optimum
cyclic behaviors. We define a location (cluster) as prevalent for a
specific time slot if the cumulated duration of association in this
location is greater than in the others locations visited during the
same time slot. For a specific time slot we take into account only
the nodes which have been associated the preceding week of
activity in the same time slot. Finally, each day are defined
through 24 time slots of one hour each.

\begin{figure}
 \includegraphics[width=8cm, height=6cm]{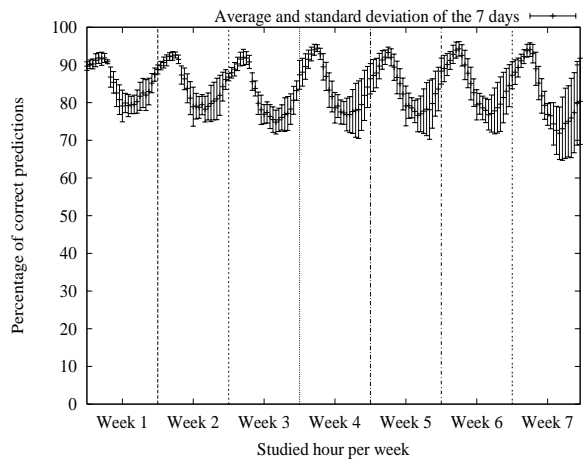}
 \caption{Percentage of nodes which have been associated to the same
prevalent location of the week before during the same time slot.}
 \label{fig:average_days}
\end{figure}

In the Fig.~\ref{fig:average_days}, we analyze the persistence of
the prevalence of a predicted location for each time slot by making
the distinction between the days. This persistence is analyzed, each
time, through two consecutive weeks of activity. We plot the
percentage of nodes for which the analysis of the time slots
have been possible.

From this figure we can make three main observations: (i) from weeks
to weeks, the percentage of correct predictions stays approximately
high and stable between 75 and 95\%. (ii) The differences between
the days are not strong and denote that making the distinction
between the days does not affect the accuracy of the predictions.
(iii) The percentage of nodes which have correct predictions
is minimal in diurnal periods between (in average 75\%) and maximal
the nights (in average 90\%). This is because we have more observed
nodes during diurnal periods than in nights. The increasing
number of nodes and their high activity in diurnal periods
decreases sensibly the results.

However, this figure does not give information on the correctness
prediction of the intra-day sequence of prevalent locations as we
consider each time slot independently. We thus consider that through
the scheme 1 (the expanding of location prevalence in the following
empty slot) and changes in pattern of mobility, that the accuracy of
the predictions will slightly decrease.

\subsubsection{Persistent periods of activity}
\label{sssec:periods_activity}

As mentioned in Section~\ref{subsec:nodecentric}, the period of
activity is a node-related parameter. In this subsection,
we analyze the differences among the nodes and the persistence
of this parameter.

In the following, we provide results about nodes which present
a certain consistency in their activity in the network. For
instance, results about mobility patterns every Monday are based on
the nodes who were active the four Monday of the observed
period. In the same way, weekly results are based on nodes who
were active each day of at least one week. The constraint about the
consistency in the activity is mostly motivated by the necessity to
compare patterns of mobility on a same plan, and to be able to
analyze persistence in behaviors.

\begin{figure}
\includegraphics[width=4.1cm, height=5cm]{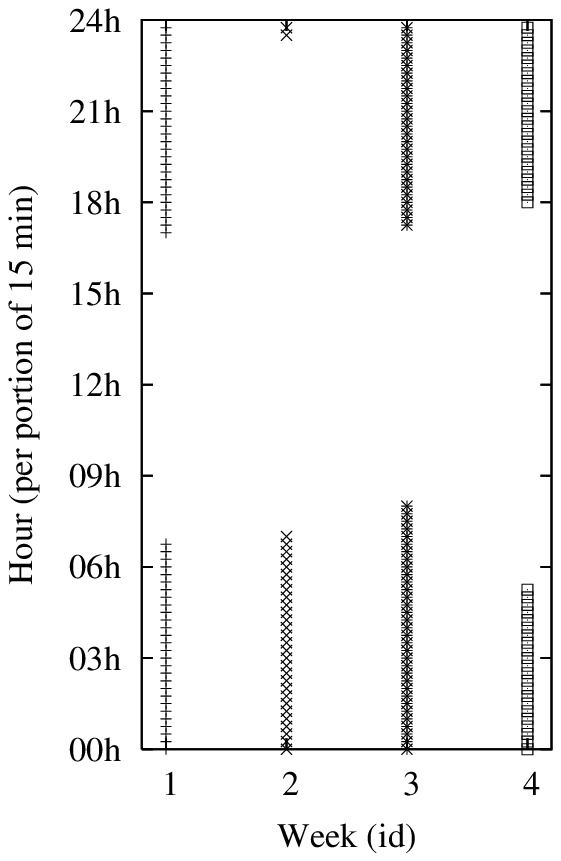}
\includegraphics[width=4.1cm, height=5cm]{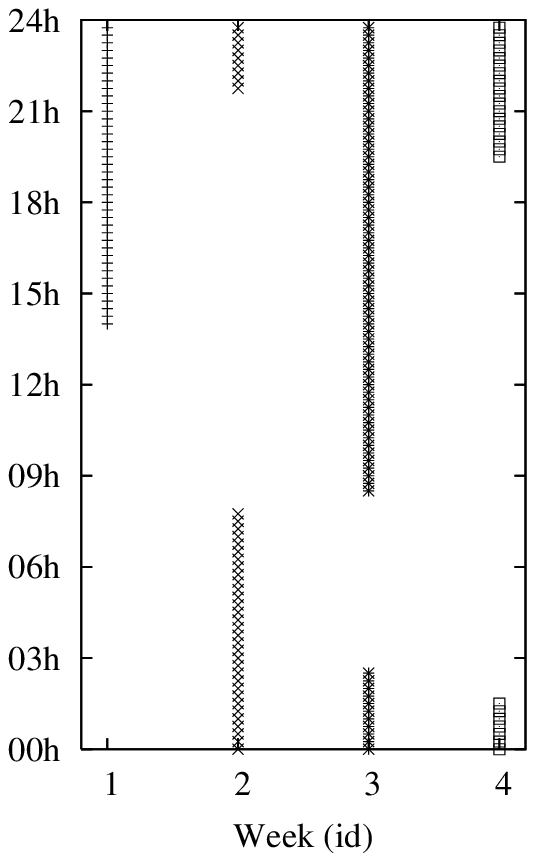}
\caption{Different types of periods of activity. It shows the
results for two nodes on Monday of each week of our observed
period.} \label{fig:period_pattern}
\end{figure}

The Fig.~\ref{fig:period_pattern} presents the periods of activity
of two different nodes for the same day (Monday) of the four
weeks. In this figure, we analyze the behavior of the nodes
within one particular day. While for the node at the left we
can observe regularity in the period of non activity (between 9h and
17h), the periods of activity of the node at the right are
completely different. The main important aspect is in the regularity
and so the persistence of the period of activity of the node
at the left. For the node at the right, the cumulated periods
of activity represents approximately an activity all the day. This
is an entirely different behavior which let suppose that the
node can have access to the network at any time.

If these two patterns of activity can be classified in two different
categories, it should exist two more different categories in this
classification: (i) nodes active along the day and (ii)
nodes active only during work hours. Under this
classification, it is different ways for different needs which can
dictate the behavior of the nodes in the access of the
network. This supposes persistence at short and average terms of the
behaviors. However, we can not conclude on this intra-day behavior
without comparing with the periods of activity the others days of
the week.

\begin{figure}
\includegraphics[width=8cm, height=6cm]{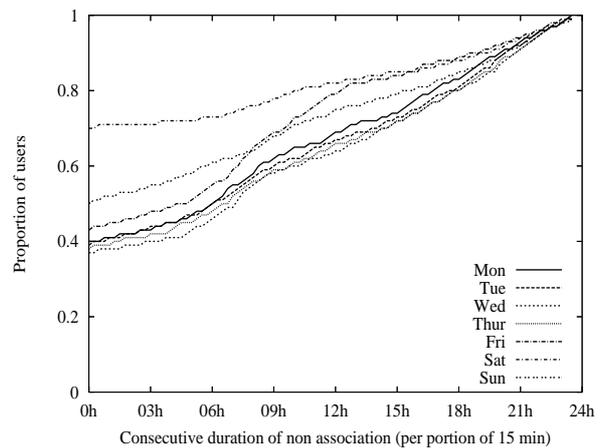}
\caption{For each day, the CDF of the length of time slots where the
nodes have never been active.} \label{fig:day_hole}
\end{figure}

The Fig.~\ref{fig:day_hole} represents, for each day, the
\textsc{CDF} of the ma\-ximum consecutive duration of non activity
in the network. For each day, the periods of activity of the four
weeks are cumulated to give the periods of non activity. Then the
durations of non activity observed correspond to periods where the
nodes have never been active in the network. The first
observation is that around 40\% of nodes are active all the
day for every week days and 48\% to 70\% the week-end. The second
observation is that for 60\% of nodes the maximum consecutive
period of non activity is nearly the same for each day of the week
days except for the Fridays. 40\% of these nodes is, for 12
hours and less, absent from the network.

It is important to note that the durations of non activity are
clearly different between the week days and the week-end. The
nodes which are subject to constraints (social or not) the
week days are more free to access the network differently the
week-end.

To summarize, the nodes can have cyclic pe\-riods of activity
which can be for a majority of them persistent. These behaviors are
not directly correlated from one day to another. It is thus required
to take each day independently.

\section{Evaluation}
\label{sec:evaluation}

As explained previously, a locator is assigned to each time slot. To
generate an agenda, each mobile node first determines for each time
slot what is the prevalent cluster. At the end of the validity period 
of its agenda, it schedules the new locators' positions. For each 
time slot, the locator is positioned in the prevalent cluster. 
To evaluate the accuracy of our predicted agenda, we need
to check for each time slot whether the mobile node was close to
its locator. Therefore we verify if the mobile node has visited its
locator's cluster during the given time period.

\subsection{Methodology}

For this evaluation we enlarge the period of analysis. It starts now
from the 6th of January 2003 (timestamp 1041829200) to finish the
end of the 29th of February 2004 (timestamp 1078117200). We choose
this period length to see how the learning step can improve the
accuracy of our predictions after each school holidays and with the
arrival of new nodes.

To evaluate the pertinence of our agenda, we check whether the
mobile node has visited its locator's cluster during a time slot.
If we find a match, we consider the mobile node is effectively
close to its locator. We perform this comparison for each time
slot, each time a mobile node was connected to the network at least
once during a time slot. We do not consider time slots where a node
is inactive. Similarly, we do not include in our study the
reliability of the first bootstrapped agenda, as it does not
reflect an observed mobility pattern. The creation of the first
agenda is detailed in Section~\ref{subsubsec:bootstrap}.

For each time slot, we can thus define a good or a bad prediction.
We  evaluate the accuracy of the agenda as the ratio of bad
predictions ($NbBadPredictions$) over the total number of
considered time slot  ($TotalNbSlots$). We define the error of
prediction in the agenda as $A_{er}$ by:

\[
A_{er} = \frac{NbBadPredictions * 100}{TotalNbSlots}.
\]

W.l.g, we choose to update, each week and for each node the 
clusters by taking into account the recent patterns of mobility
but also all the preceding history of patterns of mobility. The 
update of the clusters is made after the agenda evaluation and 
we make sure that the preceding locator predictions are still 
correct even if the eligible locators for the clusters have changed.

\subsection{Evaluation of the $A_{er}$}
\begin{figure}
\includegraphics[width=8cm, height=6cm]{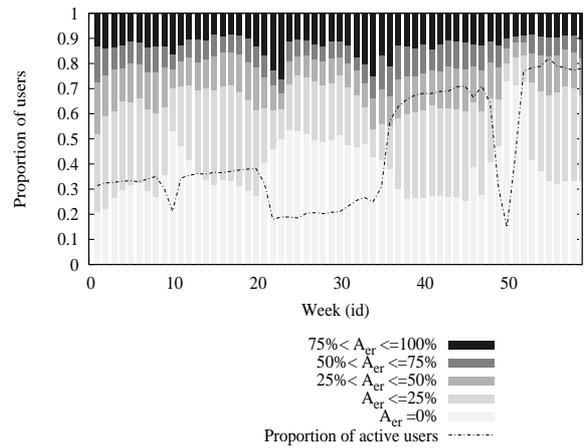}
\caption{Histogram of the proportion of nodes according to their
$A_{er}$ for each week.} \label{fig:presence}
\end{figure}

For each week in Fig.~\ref{fig:presence}, we plot the histogram 
of the proportion of nodes which have an $A_{er}$ value of
0\% (perfect prediction), less or equal to 25\%, 50\%, 75\% and
100\%.  We used a minimum of one week and up to two weeks of
mobility history to predict the value of the agenda.

The first observation is the large proportion of nodes having an
accurate agenda for the all week, $A_{er}=0\%$. Peaks of this
phenomenon can be observed at summer breaks or during Christmas
holiday. This certainly results from a static behavior of nodes.
This proportion increases as the number of active node decreases.

We also observed that only around 40\% of the average proportion 
of nodes have at least 25\% of incorrect predictions in their agenda.
Depending on the observed week, this  proportion of nodes varies
between 20 and 50\%. An average of 15\% of the proportion of nodes 
have more than 75\% of wrong predictions in their agenda and 
approximately 10\% have between 50\% and 75\% of bad accuracy.

It is important to note that if a node is not active during a time
slot, a default locator is set in the agenda, thus repeating the
value of the previous locator as explained in
Section~\ref{subsubsec:description}. A bad prediction can be easily
explained by a lack of information about a node mobility pattern.
This phenomenon can be observed in the Fig.~\ref{fig:presence}, when
the proportion of active nodes increases and remains stable
(marked as a line on the plot). We can note that the persistence of
the patterns of mobility then improves the accuracy of the
generated agenda. This accuracy decreases logically at the beginning
and at the end of the different holidays periods as mobility
constraints are relaxed during holidays and new nodes join the
network.

In facts, the reasons of most of the wrong matches are simple: (i)
often, nodes active only two weeks in the observed period do not
develop any cyclic pattern of mobility and thus have an high
$A_{er}$ value. (ii) More rarely, places which certainly constitute 
part of a path between two significant areas are assigned to time slots
(through scheme 1). Finally, (iii) overlapping between prevalent
places for one node occurs most of the time between 9am and 3pm.

These first results appear extremely promising for \otiy, our
location scheme, as it give in much cases an agenda that can allow
the optimization of  the localization process.

\subsection{Impact of the history length}
\begin{figure}
\includegraphics[width=8cm, height=6cm]{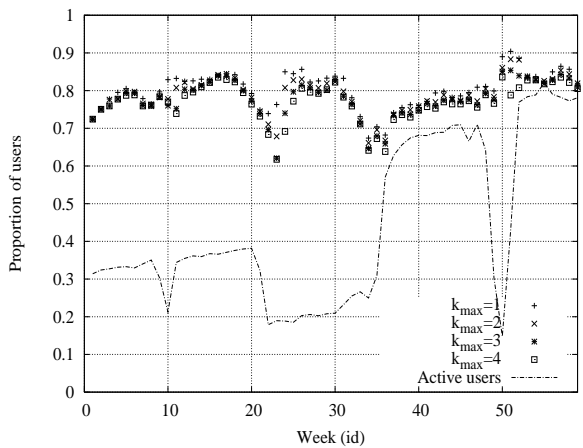}
\caption{Impact of history length used to generate the agenda on the
number of correct matches ($A_{er}\leq 50\%$).}
\label{fig:history_length}
\end{figure}

In the Fig.~\ref{fig:history_length}, we analyze the contribution of
the length of mobility history used to create the agenda on the
percentage of correct matches. Here, we plotted only the case where
there are up to 50\% bad matches in the agenda prediction.

We get the higher proportion of nodes having at least 50\% of the
time a good estimation of their agenda prediction  for a mobility
history of one week. This means that recent mobility patterns are
sufficient to estimate an accurate agenda. Nevertheless, the
duration  of activity during the current and preceding weeks also
has a clear impact on the $A_{er}$.

A convergence between different length of history can be observed
after as a result of more stable behavior of nodes. But using long
mobility history, one can note that the agendas appears more
sensitive to changes of mobility patterns. On the figure (week 11,
weeks 20 to 23, and weeks 32 to 35), we clearly see that the
accuracy  of the agendas drop roughly at the beginning and the end
of the  school holidays period.

In order to get an agenda reactive to the changes of mobility
pattern  but also be able to gather enough data about nodes'
behavior to get  the right locator estimate, we choose in the rest
of our tests to use  at a two weeks mobility history.

\subsection{Comparison between the different days}
\begin{figure}
\includegraphics[width=8cm, height=6cm]{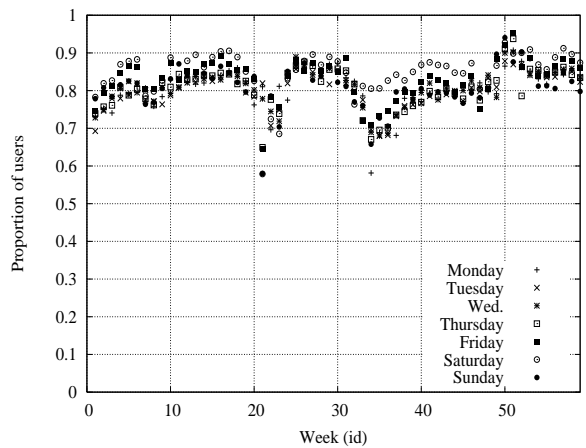}
\caption{Proportion of the nodes which have an $A_{er}\leq 50\%$ for
each day.} \label{fig:compare_days}
\end{figure}

It is interesting to check whether  the mobility behaviors affect
differently the percentage of incorrect predictions depending
particular week days. In the Fig.~\ref{fig:compare_days} we compare
the proportion of nodes which have an $A_{er}\leq50\%$ for
each day of the week.

The first observation, is that the proportion of nodes with
the  same accuracy in their agenda remains approximately the same
whatever day is considered (between 70  and 90\%). However, the
variations of proportion are more pronounced  for week-ends and
Mondays. These days are more sensitive to changes in nodes' behavior
at the beginning and at the  end of the holidays. The rest of the
time the pertinence of the agenda seems more  important for these
three days.

We envision taking advantage of  this phenomenon to better detect
changes in behaviors and to more accurately restart the
bootstrapping procedures.

\subsection{Evaluation of the physical distance}
\begin{figure}
\includegraphics[width=8cm, height=6cm]{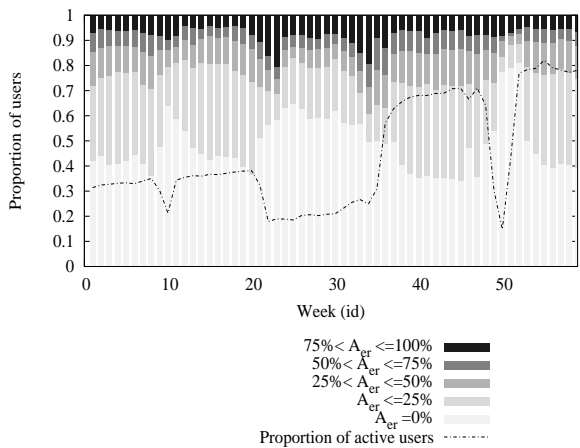}
\caption{Histogram of the proportion of nodes according to their
$A_{er}$ values for each week. In this figure, we consider also a correct
prediction if the node has visited at least one cluster connected
(at one hop) to the locator.} \label{fig:compare_onehop}
\end{figure}

As mentioned before, the topology of the network is not provided
with the data traces. As a consequence we can not determine the
topological path length(s) between the visited areas during a time
slots and the locator's position.

In order to have an idea of how close the nodes have been from the 
locator' cluster, we use the graph from which is built the clusters. 
In the Fig.~\ref {fig:compare_onehop}, we evaluate $A_{er}$ considering 
a prediction as correct, if during a time slot, a node has visited the 
locator' cluster until a cluster one hop away.

Comparing the results with Fig.~\ref{fig:presence}, we observe that
the number of nodes with an $A_{er}$ value of up to 50\% of
their agenda increases roughly by 10\% for each week. The number of
nodes  which have made more than 75\% of incorrect predictions
is reduced to, in average, 5\% per week.

These results show, that for the majority of the nodes and
most  of the time we successfully provide a locator close to the
current physical location of the nodes. We therefore can provide an
agenda that satisfactory reflects nodes  mobility.

\section{Related Work}
\label{sec:relatedwork}

The problematic of reducing the amount of generated location updates
in a wireless network has been well studied in \textsc{PCS} and is
very close to our work. Even if the proposals are not really adapted
to IEEE~802.11 wireless mesh networks, we will relate the different
concepts of the most relevant work.

Tabbane is one of the firsts to have introduced the node (mobility)
profiling to improve location management in \textsc{PCS}.
In~\cite{tabbane} the profiling is operated by the network and
shared with the node's subscriber identity module \textsc{SIM}.
Thanks to this profiling, whatever the period of time $[t_i,t_j)$
the system can find a list of areas where the nodes could be. This
list of areas is decreasingly ordered by the probabilities of being
in the different areas. One probability is dependent of a function
associated with and has several parameters such as the time, the
pattern of mobility, the last location, the weather etc. Until the
node is in one of these areas it does not update its location. When
the system needs to locate him, it asks sequentially the different
areas within the list. Two notions are shared with our approach: (i)
the node profiling although in \otiy\ it is the nodes which make
their self-profiling. (ii) The relation with the time. However, in
Otiy the period of time are predefined and only one area (anchor) is
assigned to each time slot.

Chuon \textit{et al.} in \cite{chuon} by calculating the prevalence
of the daily different visited cells create an \textit{node
profile graph} (\textsc{IPG}) per node and through their monitoring
by the network. This graph is composed by vertices (cells) where the
normalized probability of visit on N-days is greater than a
specified value (comprised between $0$ and $1$) and the connectors
between these cells (also called anchors). Furthermore, the vertices
in the \textsc{IPG} are classified by decreasing order of
probability of visit. Until the node stays in the set of vertices in
his \textsc{IPG} it does not have to updates its location. To locate
it in its \textsc{IPG}, the network pages it in the decreasing
order of probability of the anchors as in~\cite{pollini}. We differ
from this approach in several ways. We do not cumulate daily
patterns of mobility to see a prevalent mobility graph. Rather, we
distinguish each daily pattern of mobility of the week to capture
the different constraints and interests which are dependent of the
days. Although they used the division in time slots to prove that
their \textsc{IPG} captures well the diurnal mobility of the nodes,
they do not used these time slots to make the relationship, that we
make, between a particular area and the time. From this point, our
proposal is clearly different of the methods used to updates the
locations and to page the nodes.

In \cite{wu01personal}, Wu \textit{et al.} mine the mobility
behavior nodely (operated by the nodes) from long term
mobility history. From this information they evaluate the
time-varying probability of the different nodely-defined
regions. The prevalence of a region on the time is defined through a
cost model. Finally, they obtain a vector $<time,area>$ of mobility
which will define the region to be paged in function of the time.
The location updates and paging schemes are approximately the same
than the afore-presented proposals. Here, the prevalence of a
particular area on the time is more flexible and accurate than ours.
However, the complexity of the algorithm is more important.
Furthermore, the length of the vector of mobility can be a lot more
important than a division in time slots of equal duration.

Finally in IEEE~802.11 wireless networks, Ghosh \textit{et al.} have
profiling the nodes associations sequences by making the difference
between the days in~\cite{ghosh_profiling}. The result is several
patterns of mobility such as \textit{Weekend Profile}, \textit{Home
Profile}, etc. Here, the areas are called ``sociological hub'' and
are defined at a building granularity. With this set of profiles
they are capable to determine which mobility profile the node follow
currently (from the firsts associations) and/or define a window of
day with attributed mobility profiles. This approach is sensibly
different of Otiy but here, we recognize the necessity to
differentiate the patterns of mobility which could be different from
day to day. Furthermore, our significant areas (they ``hub'') are
determined by the local micro-mobility of each node and are
not based on any geographical information (e.g., building
segmentation).

\section{Conclusion}
\label{sec:conclusion}

We introduced Otiy, a node-centric architecture to control the
propagation of the location updates in networks of mobile nodes. To
limit the impact of location updates, each node creates an
agenda of locators, which are access points at which the node will
probably get associated with according to the past and present
patterns of mobility. The idea behind this approach is to make
location update overhead more localized.

The keys principles we achieve are:

\begin{itemize}

\item A fully decentralized location service.

\item Short path lengths for location updates.

\item Shared knowledge on absolute time locations. At anytime, both
the node and its interlocutor(s) know the location of the
locator.

\end{itemize}

To achieve our goals, we proposed to gather nearby APs in terms of
node roaming events in order to create areas (clusters) of
micro-mobility. Our clustering algorithm does not use any
geographical information and is based only on the sequence of
associations/dis\-associa\-tions. This drastically limits the 
propagation of updates due to artifacts such as ping-pong effects.

Using around one year of mobility traces, we have in average 75\% 
of the nodes that have at least 50\% of their time slots which match
exactly the right predictions. These results encourage us to keep
improving \otiy\ and evaluate it using real testbeds.

\bibliographystyle{abbrv}
\bibliography{bib/boc_otiy.bbl}

\end{document}